\documentclass[reprint,superscriptaddress,amsmath, amssymb,aps,prd,notitlepage,longbibliography,floatfix,nofootinbib,twocolumn]{revtex4-1}

\usepackage{subfigure}
\usepackage{amsmath}
\usepackage{lipsum}
\allowdisplaybreaks[4]
\usepackage{cancel}
\usepackage{extarrows}
\usepackage{tensor}     
\usepackage{float}
\usepackage[final]{graphicx}   
\usepackage[
colorlinks=true,        
citecolor=blue,         
linkcolor=blue,         
urlcolor=blue           
]{hyperref}             
\usepackage{bm}         
\usepackage{xcolor}     
\usepackage{lipsum}
\usepackage{color}      
\usepackage[utf8]{inputenc} 
\usepackage[section]{placeins} 
\usepackage{appendix}
\usepackage{units}
\usepackage[capitalise]{cleveref}
\usepackage{units}
\usepackage{adjustbox}
\usepackage{makecell}
\newcommand{\nc}{\newcommand*}

\nc{\xbar}{\bar{x}}
\nc{\rhoeq}{\rho_{\mathrm{eq}}}
\nc{\zeq}{z_{\mathrm{eq}}}
\nc{\tla}{\tilde{\lambda}}
\nc{\bt}{\beta}
\nc{\dt}{\delta}
\nc{\Dt}{\Delta}
\nc{\vj}{\vec{j}}
\nc{\vl}{\vec{l}}
\nc{\hx}{\hat{x}}
\nc{\hy}{\hat{y}}
\nc{\bj}{\bm{j}}
\nc{\mJ}{\mathcal{J}}
\nc{\mP}{\mathcal{P}}
\nc{\Msun}{M_\odot}
\nc{\app}{\approx}
\nc{\av}[1]{\langle #1 \rangle}
\nc{\eq}[1]{Eq.~\eqref{#1}}
\nc{\al}{\alpha}
\nc{\Xstar}{X_{\ast}}
\nc{\fpbh}{f_{\mathrm{pbh}}}
\nc{\vth}{\vec{\theta}}
\nc{\vla}{\vec{\lambda}}
\nc{\vd}{\vec{d}}
\nc{\Mmin}{M_{\mathrm{min}}}
\nc{\rmd}{\mathrm{d}}
\nc{\mmin}{{m_{\mathrm{min}}}}
\nc{\mmax}{{m_{\mathrm{max}}}}
\nc{\mR}{\mathcal{R}}
\nc{\tmR}{\tilde{\mathcal{R}}}
\nc{\s}{\sigma}
\nc{\ogw}{\Omega_{\mathrm{GW}}}
\nc{\addref}{[\textcolor{red}{add ref}] }
\nc{\Om}{\Omega}
\nc{\gm}{\gamma}
\nc{\Gm}{\Gamma}
\nc{\gpcyr}{\mathrm{Gpc}^{-3}\,\mathrm{yr}^{-1}}
\nc{\Eq}[1]{Eq.~\eqref{#1}}
\nc{\Fig}[1]{Fig.~\ref{#1}}
\nc{\Table}[1]{Table~\ref{#1}}
\nc{\lvc}{LIGO/Virgo} 
\nc{\Sec}[1]{Sec.~\ref{#1}}
\nc{\eg}{\textit{e.g.~}}
\nc{\SNR}{\mathrm{SNR}}
\nc{\be}{\mathbf{\epsilon}}
\nc{\bn}{\mathbf{n}}
\nc{\bd}{\mathbf{d}}
\nc{\ba}{\mathbf{a}}
\nc{\eps}{\epsilon}
\nc{\bnu}{\mathbf{\nu}}
\nc{\mb}{\mathbf}
\nc{\bbt}{\mathbf{t}}
\nc{\bth}{\mathbf{\theta}}
\nc{\bep}{\mathbf{\epsilon}}
\nc{\uni}{\mathrm{U}}
\nc{\logu}{\operatorname{\mathrm{log-U}}}
\nc{\RN}{\mathrm{RN}}
\nc{\BN}{\mathrm{BN}}
\nc{\GN}{\mathrm{GN}}
\nc{\mcN}{\mathcal{N}}
\nc{\GWB}{\mathrm{GW}}
\nc{\yr}{\mathrm{yr}}
\nc{\Am}{\mathcal{A}}
\nc{\Dm}{\mathcal{D}}
\nc{\Hm}{\mathcal{H}}
\nc{\sovast}{Soviet Ast.}

\nc{\kmsmpc}{\mathrm{km\ s^{-1} Mpc^{-1}}}
\nc{\lcdm}{\Lambda\mathrm{CDM}}
\nc{\ev}{\mathrm{eV}}

\nc{\mrm}{\mathrm}
\nc{\BE}{B\scriptsize{AYES}\normalsize{E}\scriptsize{PHEM}\normalsize  }

\nc{\Ostgw}{\Omega_{\mathrm{GW}}^{\mathrm{ST}}}
\nc{\Ottgw}{\Omega_{\mathrm{GW}}^{\mathrm{TT}}}
\nc{\Ovlgw}{\Omega_{\mathrm{GW}}^{\mathrm{VL}}}
\nc{\Oslgw}{\Omega_{\mathrm{GW}}^{\mathrm{SL}}}
\nc{\cosxi}{\beta}

\nc{\gmPL}{\gamma_{\mathrm{PL}}}
\nc{\APL}{A_{\mathrm{PL}}}

\def\({\left(}
\def\){\right)}
\def\[{\left[}
\def\]{\right]}

\def\e{\begin{equation}}
\def\q{\end{equation}}
\def\m{\begin{eqnarray}}
\def\n{\end{eqnarray}}
\nc{\red}[1]{\textcolor{red}{#1}}

\begin{document}

\title{The Impact of the Hubble Tension on the Evidence for Dynamical Dark Energy}

\author{Ye-Huang Pang}
\affiliation{School of Fundamental Physics and Mathematical Sciences, Hangzhou Institute for Advanced Study, UCAS, Hangzhou 310024, China}
\affiliation{School of Physical Sciences,
    University of Chinese Academy of Sciences,
    No. 19A Yuquan Road, Beijing 100049, China}
\affiliation{CAS Key Laboratory of Theoretical Physics,
    Institute of Theoretical Physics, Chinese Academy of Sciences,Beijing 100190, China}
\author{Xue Zhang}
\email{corresponding author: zhangxue@yzu.edu.cn}
\affiliation{Center for Gravitation and Cosmology,
    College of Physical Science and Technology,
    Yangzhou University, Yangzhou 225009, China}
\author{Qing-Guo Huang}
\email{corresponding author: huangqg@itp.ac.cn}
\affiliation{School of Fundamental Physics and Mathematical Sciences, Hangzhou Institute for Advanced Study, UCAS, Hangzhou 310024, China}
\affiliation{School of Physical Sciences,
    University of Chinese Academy of Sciences,
    No. 19A Yuquan Road, Beijing 100049, China}
\affiliation{CAS Key Laboratory of Theoretical Physics,
    Institute of Theoretical Physics, Chinese Academy of Sciences,Beijing 100190, China}


\begin{abstract}
Recent findings from the Dark Energy Spectroscopic Instrument (DESI) Data Release 2 (DR2) favor a dynamical dark energy characterized by a phantom crossing feature. This result also implies a lower value of the Hubble constant, thereby intensifying the so-called Hubble tension. To alleviate the Hubble tension, we consider the early dark energy and explore its impact on the evidence for dynamical dark energy including the Hubble constant calibrated by the SH0ES collaboration. 
We find that incorporating SH0ES prior with CMB, DESI DR2 BAO and Pantheon Plus/Union3/DESY5 data reduces the preference to dynamical dark energy to $1.5\sigma/1.4\sigma/2.4\sigma$ level, respectively. Our results suggest a potential tension between the Hubble constant $H_0$ of the SH0ES measurement and the phantom-to-quintessence transition in dark energy favored by DESI DR2 BAO data.

\end{abstract}
\maketitle

\section{Introduction}

The recent measurements of baryon acoustic oscillations (BAO) from the Dark Energy Spectroscopic Instrument (DESI) second data release (DR2) \cite{DESI:2025zpo, DESI:2025zgx} has significantly enhanced the precision of cosmological inferences, providing stronger evidence for dynamical dark energy compared to its first data release (DR1). Specifically, the combination of Cosmic Microwave Background (CMB) data, BAO measurements from DESI DR2 and supernova data from DESY5 indicates a $4.2\sigma$ preference for dynamical dark energy within $w_0w_a$CDM model. In this model, the dark energy equation of state $w(a)$ is parameterized as 
\begin{equation}
    w(a) = w_0 + w_a(1-a), 
\end{equation}
where $a$ represents the scale factor \cite{Chevallier:2000qy, Linder:2002et}. Additionally, the combination of CMB+DESI DR2 BAO+Pantheon Plus/Union3 datasets yields dynamical dark energy at $2.8\sigma/3.8\sigma$ confidence levels (C.L.), respectively. Further investigations, as detailed in \cite{Lodha:2025qbg}, explore the dynamical characteristics of dark energy through various methodologies, consistently supporting the findings within the $w_0w_a$CDM model. For constraints on alternative dark energy models using either DESI DR1 or DR2 data, we refer readers to, e.g. \cite{Wang:2024hks, Cortes:2024lgw, Carloni:2024zpl, Berghaus:2024kra, Giare:2024smz, Montani:2024pou, Wang:2024dka, Yang:2024kdo, Shlivko:2024llw, Huang:2024qno, Wang:2024rus, Dinda:2024kjf, DESI:2024kob, Bhattacharya:2024hep, Ramadan:2024kmn, Mukherjee:2024ryz, Wang:2024hwd, Gialamas:2024lyw, Notari:2024rti, Chudaykin:2024gol, Liu:2024gfy, Orchard:2024bve, Hernandez-Almada:2024ost, Mukhopadhyay:2024fch, Li:2024qso, Ye:2024ywg, Giare:2024gpk, Pang:2024qyh, Wolf:2024eph, Taule:2024bot, Giare:2024ocw, Li:2024qus, Payeur:2024dnq, Gao:2024ily, DESI:2025hce, Anchordoqui:2025fgz, Pan:2025psn, Ishiyama:2025bbd}.

Furthermore, the constrains on the $w_0w_a$CDM model indicate an evolutionary behavior in the dark energy equation of state, transitioning from $w<-1$ (phantom-like \cite{Caldwell:1999ew}) at high redshifts to $w>-1$ (quintessence-like \cite{Ratra:1987rm, Ferreira:1997hj, Wetterich:1987fm, Scherrer:2007pu}) at low redshifts. This behavior implies that the dark energy density decreases with the expansion of the Universe at low redshifts, ultimately resulting in a lower value for the Hubble constant $H_0$.
As reported in \cite{DESI:2025zgx}, the Hubble constant is determined to be $H_0 = 63.6^{+1.6}_{-2.1} \; \kmsmpc$ in the $w_0w_a$CDM model, constrained by CMB+DESI DR2 BAO. This value exhibits a tension of $5.3\sigma$ with the SH0ES measurement ($H_0 = 73.17 \pm 0.86$ $\kmsmpc$) \cite{Breuval:2024lsv}. Furthermore, when considering data combinations such as CMB+DESI DR2 BAO+Pantheon Plus/Union3/DESY5, the corresponding constraints are found to be $H_0 = 67.51 \pm 0.59$, $65.91 \pm 0.84$, and $66.74 \pm 0.56 \; \kmsmpc$. These values indicate $5.4\sigma$, $6.0\sigma$, and $6.3\sigma$ tension with respect to the SH0ES measurement, respectively.
Several of these results further intensify the existing tension of $5.6\sigma$ between the Planck 2018 results \cite{Planck:2018vyg} ($H_0 = 67.27 \pm 0.60\ \kmsmpc$) and those from SH0ES measurements. 
The Hubble tension has been extensively investigated in recent years. For comprehensive reviews on this topic, we refer readers to works by \cite{Verde:2019ivm, DiValentino:2021izs, Abdalla:2022yfr}.
Given that there is a preference for dynamical dark energy indicated by DESI DR2, modified cosmological models that favor higher values of $H_0$ may be incompatible with the late-time dynamical dark energy. This potential contradiction could weaken the statistical preference for such dynamical dark energy scenarios.

This interplay motivates a critical investigation: whether the evidence for dynamical dark energy remains robust when simultaneously addressing the Hubble tension. In this article, we incorporate the early dark energy (EDE) model \cite{Poulin:2018dzj, Poulin:2018cxd, Lin:2019qug, Niedermann:2019olb, Ye:2020btb, Karwal:2021vpk, Poulin:2023lkg} as a representative higher-$H_0$ scenario, with a specific focus on the $n=3$ axion-like EDE framework.

\section{Evidence for dynamical dark energy in the $w_0w_a$CDM+EDE model}
\begin{table*}[!htbp]
    \centering
    \scalebox{0.95}{\begin{tabular} { l|  c|  c|   c|   c}
        \hline
          Parameters
         & \makecell{CMB+DESI DR2 BAO\\+SH0ES}
         & \makecell{CMB+DESI DR2 BAO\\+Pantheon Plus+SH0ES}
         & \makecell{CMB+DESI DR2 BAO\\+Union3+SH0ES}
         & \makecell{CMB+DESI DR2 BAO\\+DESY5+SH0ES}
         \\
        
        
        
        
        
        
        
        
        
        \hline
        {$\log(10^{10} A_\mathrm{s})$}
         & $3.071(3.072)\pm 0.013$
         & $3.069(3.067)\pm 0.013$
         & $3.069(3.071)\pm 0.013$
         & $3.068(3.066)\pm 0.013$
         \\
        {$n_\mathrm{s}$}
         & $0.9891(0.9923)\pm 0.0060$
         & $0.9885(0.9889)\pm 0.0061$
         & $0.9888(0.9848)\pm 0.0057$
         & $0.9881(0.9890)\pm 0.0057$
         \\
        {$\Omega_\mathrm{b} h^2$}
         & $0.02261(0.02253)\pm 0.00019$
         & $0.02259(0.02248)\pm 0.00020$
         & $0.02259(0.02246)\pm 0.00019$
         & $0.02258(0.02256)\pm 0.00019$
         \\
        {$\Omega_\mathrm{c} h^2$}
         & $0.1293(0.1330)\pm 0.0028$
         & $0.1297(0.1310)\pm 0.0029$
         & $0.1297(0.1295)\pm 0.0027$
         & $0.1299(0.1303)\pm 0.0028$
         \\
        {$\tau_\mathrm{reio}$}
         & $0.0603(0.0578)^{+0.0066}_{-0.0074}$
         & $0.0591(0.0563)\pm 0.0071$
         & $0.0593(0.0575)\pm 0.0070$
         & $0.0584(0.0568)\pm 0.0070$
         \\
        {$H_0 [\kmsmpc]$}
         & $71.82(72.45)\pm 0.74$
         & $71.74(72.02)\pm 0.77$
         & $71.77(71.77)\pm 0.69$
         & $71.65(71.86)\pm 0.72$
         \\
        {$f_{\mathrm{EDE}}$}
         & $0.115(0.139)^{+0.022}_{-0.019}$
         & $0.115(0.124)^{+0.023}_{-0.019}$
         & $0.115(0.112)^{+0.022}_{-0.018}$
         & $0.115(0.120)^{+0.022}_{-0.020}$
         \\
        {$\log_{10}a_c$}
         & $-3.63(-3.58)^{+0.12}_{-0.072}$
         & $-3.62(-3.57)^{+0.12}_{-0.072}$
         & $-3.62(-3.56)^{+0.11}_{-0.067}$
         & $-3.62(-3.61)^{+0.13}_{-0.066}$
         \\
        \hline
        
        {$\chi^2_{\mathrm{bestfit}}$}
        & 11005.10
        & 12411.71
        & 11033.18
        & 12656.20\\ 
        \hline
        \end{tabular}}
    \caption{The mean (best-fit) value and $1\sigma$ C.L. of the cosmological parameter and the corresponding $\chi^2$ for best-fit values in $\Lambda$CDM+EDE model.}
    \label{tab:lcdm_ede_table}
\end{table*}

\begin{table*}[!htbp]
    \centering
    \scalebox{0.95}{\begin{tabular} { l|  c|  c|   c|   c}
        \hline
          Parameters
         & \makecell{CMB+DESI DR2 BAO\\+SH0ES}
         & \makecell{CMB+DESI DR2 BAO\\+Pantheon Plus+SH0ES}
         & \makecell{CMB+DESI DR2 BAO\\+Union3+SH0ES}
         & \makecell{CMB+DESI DR2 BAO\\+DESY5+SH0ES}
         \\

        \hline
        {$\log(10^{10} A_\mathrm{s})$}
         & $3.055(3.059)\pm 0.014$
         & $3.060(3.055)\pm 0.014$
         & $3.060(3.054)\pm 0.014$
         & $3.060(3.067)\pm 0.014$
         \\
        {$n_\mathrm{s}$}
         & $0.9784(0.9845)^{+0.0072}_{-0.0083}$
         & $0.9857(0.9841)\pm 0.0070$
         & $0.9863(0.9846)\pm 0.0071$
         & $0.9875(0.9907)\pm 0.0069$
         \\
        {$\Omega_\mathrm{b} h^2$}
         & $0.02247(0.02237)^{+0.00018}_{-0.00020}$
         & $0.02256(0.02245)\pm 0.00019$
         & $0.02257(0.02236)\pm 0.00020$
         & $0.02258(0.02250)\pm 0.00020$
         \\
        {$\Omega_\mathrm{c} h^2$}
         & $0.1259(0.1292)\pm 0.0033$
         & $0.1300(0.1318)^{+0.0027}_{-0.0031}$
         & $0.1307(0.1325)\pm 0.0030$
         & $0.1314(0.1344)\pm 0.0030$
         \\
        {$\tau_\mathrm{reio}$}
         & $0.0553(0.0548)\pm 0.0072$
         & $0.0548(0.0520)\pm 0.0072$
         & $0.0542(0.0483)\pm 0.0072$
         & $0.0539(0.0539)\pm 0.0071$
         \\
        {$H_0 \kmsmpc$}
         & $72.16(72.57)\pm 0.83$
         & $71.40(71.31)\pm 0.70$
         & $71.23(71.25)\pm 0.74$
         & $71.01(71.28)\pm 0.73$
         \\
        {$f_{\mathrm{EDE}}$}
         & $0.070(0.102)\pm 0.031$
         & $0.109(0.119)\pm 0.026$
         & $0.114(0.126)^{+0.028}_{-0.025}$
         & $0.120(0.142)^{+0.027}_{-0.024}$
         \\
        {$\log_{10}a_c$}
         & $-3.61(-3.59)^{+0.20}_{-0.10}$
         & $-3.61(-3.59)^{+0.15}_{-0.063}$
         & $-3.61(-3.55)^{+0.14}_{-0.052}$
         & $-3.60(-3.59)^{+0.13}_{-0.051}$
         \\
        {$w_{0}$}
         & $-1.057(-1.03)\pm 0.094$
         & $-0.908(-0.878)\pm 0.048$
         & $-0.869(-0.844)\pm 0.069$
         & $-0.833(-0.817)\pm 0.052$
         \\
        {$w_{a}$}
         & $-0.07(-0.081)^{+0.28}_{-0.25}$
         & $-0.38(-0.47)^{+0.20}_{-0.18}$
         & $-0.49(-0.54)\pm 0.24$
         & $-0.57(-0.58)^{+0.22}_{-0.20}$
         \\
        \hline
        
        {$\chi^2_{\mathrm{bestfit}}$}
        & 11003.21
        & 12408.87
        & 11030.17
        & 12646.80\\ 
        \hline
        \end{tabular}}
    \caption{The mean (best-fit) value and $1\sigma$ C.L. of the cosmological parameter and the corresponding $\chi^2$ for best-fit values in $w_0w_a$CDM+EDE model. }
    \label{tab:w0wa_ede_table}
\end{table*}

\begin{table*}[!htbp]
    \centering
    \scalebox{1.0}{\begin{tabular} { l|  c|  c|   c|   c}
        \hline
          Parameterss
         & \makecell{CMB+DESI DR2 BAO\\+SH0ES}
         & \makecell{CMB+DESI DR2 BAO\\+Pantheon Plus+SH0ES}
         & \makecell{CMB+DESI DR2 BAO\\+Union3+SH0ES}
         & \makecell{CMB+DESI DR2 BAO\\+DESY5+SH0ES}
         \\
        \hline
        
        {$\Delta\chi^2_{\mathrm{MAP}}$}
        & $-3.18$
        & $-3.91$
        & $-3.80$
        & $-8.31$\\ 
        \hline
        {Significance}
        & $1.3\sigma$
        & $1.5\sigma$
        & $1.4\sigma$
        & $2.4\sigma$\\ 
        \hline
        \end{tabular}}
    \caption{$\Delta\chi^2_{\mathrm{MAP}}$ and statistical significance for dynamical dark energy.}
    \label{tab:chisq_cl}
\end{table*}





The axion-like EDE model with $n=3$ \cite{Poulin:2018dzj, Poulin:2018cxd} refers to a scalar field $\phi$ characterized by the potential
\begin{equation}
    V(\phi) = m^2f^2[1-\cos(\phi/f)]^3,
\end{equation}
where $m$ denotes the mass of scalar field and $f$ represents the axion decay constant. This potential enables the EDE to initially behave like a cosmological constant before the field rolls down the potential and begins oscillating around a critical epoch.

The EDE fractional energy density reaches its peaks at redshift $z_c$ with a corresponding value of $f_{\mathrm{EDE}}$. By modifying the expansion rate near recombination, EDE reduces the sound horizon at recombination $r_s(z_*)$ (and $r_d$), while preserving the tightly constrained angular scale of the sound horizon $\theta(z_*)$ as determined by CMB measurements \cite{Poulin:2018cxd, Knox:2019rjx}. This mechanism allows EDE to accommodate a higher value of $H_0$ without conflicting with CMB observations. Additionally, the initial field value $\phi_i$ influences the EDE dynamics and serves as an independent sampling parameter in our analysis.

We investigate a combined model that integrates EDE with the $w_0w_a$CDM framework, which we refer to as the $w_0w_a$CDM+EDE model. To evaluate the evidence for dynamical dark energy within the context of EDE cosmology, we also analyze the $\Lambda$CDM+EDE model, in which late-time dark energy is represented by a cosmological constant $\Lambda$. 

Both the $\Lambda$CDM+EDE and $w_0w_a$CDM+EDE models are constrained using the following datasets:
\begin{itemize}
    \item CMB
    \newline The CMB data include \textit{Planck} low-T \texttt{Commander} likelihood, low-E \texttt{SimAll} likelihood, \texttt{CamSpec} high-$\ell$ likelihood \cite{Rosenberg:2022sdy}, as well as the \textit{Planck} PR4 lensing likelihood \cite{Carron:2022eyg}, and the Atacama Cosmology Telescope (ACT) DR6 lensing likelihood \cite{ACT:2023dou}.
    \item DESI DR2 BAO
    \newline The data points utilized in this analysis are provided in Table IV of Ref. \cite{DESI:2025zgx} .
    \item Supernova
    \newline We employ one of the following supernova samples: Pantheon Plus \cite{Brout:2022vxf}, Union3 \cite{Rubin:2023ovl} or DESY5 \cite{DES:2024tys}.
    \item SH0ES
    \newline We impose a Gaussian prior of $H_0 = 73.17 \pm 0.86$ $\kmsmpc$, which is reported in Ref. \cite{Breuval:2024lsv}.
\end{itemize}
These datasets are largely consistent with those used in the original DESI DR2 analysis. Our analysis employs four combinations of datasets: CMB+DESI DR2 BAO+SH0ES and CMB+DESI DR2 BAO+Pantheon Plus/Union3/DESY5+SH0ES.

\begin{figure}[!htbp]
    \centering
    \includegraphics[width=\linewidth]{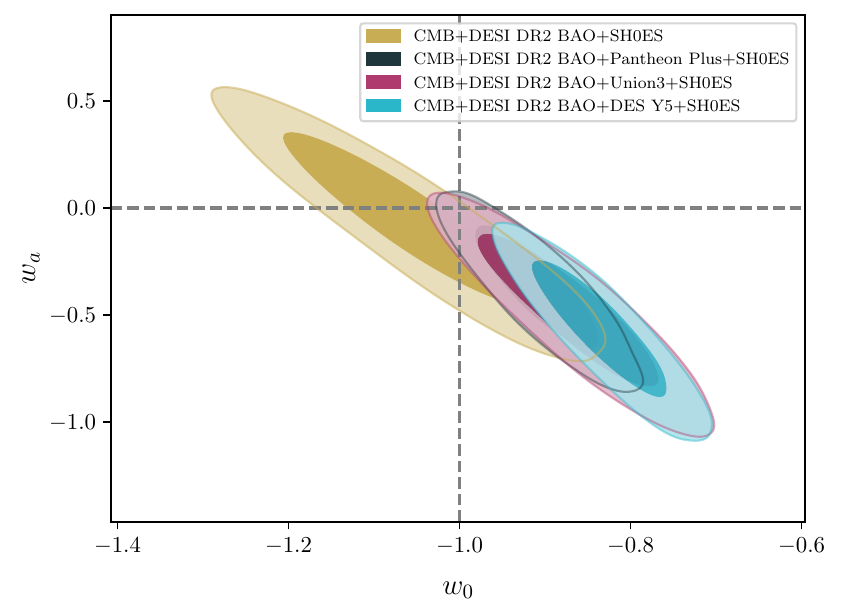}
    \caption{Marginalized posterior distributions for $w_0$ and $w_a$ in the $w_0w_a$CDM+EDE model. }
    \label{fig:w0wa}
\end{figure}

The cosmological parameters are calculated using the \texttt{AxiCLASS}\footnote{ \url{https://github.com/PoulinV/AxiCLASS}} code \cite{Blas:2011rf, Murgia:2020ryi}. 
For parameter estimation, we perform Markov Chain Monte Carlo (MCMC) sampling with the \texttt{cobaya} code \cite{Torrado:2020dgo}, ensuring convergence by applying the Gelman-Rubin diagnostic criterion \cite{Gelman:1992zz}, specifically $R-1<0.05$. 

The resulting parameter constraints are summarized in \Table{tab:lcdm_ede_table} and \Table{tab:w0wa_ede_table}. The $1\sigma$ and $2\sigma$ confidence contours for the $w_0-w_a$ are presented in \Fig{fig:w0wa}. To quantify the preference for dynamical dark energy, we calculate the $\Delta\chi^2$ between the maximum a posteriori (MAP) points of the $\Lambda$CDM+EDE and $w_0w_a$CDM+EDE models. The results, along with their statistical significance, are displayed in \Table{tab:chisq_cl}.

\section{Discussion}

As shown in Table \ref{tab:chisq_cl}, the joint analysis using CMB+DESI DR2 BAO+SH0ES yields a preference for late-time dynamical dark energy at $1.3 \sigma$ confidence level. 
Furthermore, when incorporating Pantheon Plus, Union3, or DESY5 supernova samples alongside CMB+DESI DR2 BAO+SH0ES, we observe preferences for dynamical dark energy at $1.5\sigma$, $1.4\sigma$ and $2.4\sigma$, respectively.
Compared to the constraints on $w_0w_a$CDM model derived from CMB+DESI DR2 BAO+Pantheon Plus/Union3/DESY5 data (which indicate a preference for dynamical dark energy at $2.8\sigma$, $3.8\sigma$ and $4.2\sigma$, respectively), the statistical preference for dynamical dark energy is significantly reduced.

The inclusion of SH0ES prior increases the inferred value of $H_0$,  implying a larger dark energy density at present. However, the dark energy density with the phantom-to-quintessence transition favored by the DESI DR2 BAO data decreases at low redshifts, and therefore the larger value of Hubble constant from SH0ES reduces the statistical preference for such a dynamical dark energy model. The underlying cause of this phenomenon arises from a inherent tension between higher $H_0$ measurement and the DESI DR2 BAO data.

\textit{Acknowledgements.}
QGH is supported by the grants from NSFC (Grant No.~12475065) and China Manned Space Program through its Space Application System. We acknowledge the use of HPC Cluster of ITP-CAS.

\bibliography{refs}
\end{document}